\documentclass[accepted]{uai2025} 
\usepackage{aux}
                        
\usepackage[american]{babel}
\usepackage{soul}

\AtBeginBibliography{\small}

\title{Causal Inference amid Missingness-Specific Independencies\\ and Mechanism Shifts}
\author[1,2]{\href{mailto:<johanmd@uio.no>?Subject=Your UAI 2025 paper}{Johan~de~Aguas}{}}
\author[3]{Leonard~Henckel}
\author[1,4]{Johan~Pensar}
\author[2]{Guido~Biele}
\affil[1]{%
    Department of Mathematics, University of Oslo
}
\affil[2]{%
    Department of Child Health and Development, Norwegian Institute of Public Health
}
\affil[3]{%
    School of Mathematics and Statistics. University College Dublin
}
\affil[4]{%
     INTEGREAT -- The Norwegian Center for Knowledge-Driven Machine Learning
}

\begin{document}
\maketitle

\begin{abstract}
The recovery of causal effects in structural models with missing data often relies on $m$-graphs, which assume that missingness mechanisms do not directly influence substantive variables. Yet, in many real-world settings, missing data can alter decision-making processes, as the absence of key information may affect downstream actions and states. To overcome this limitation, we introduce $lm$-SCMs and $lm$-graphs, which extend $m$-graphs by integrating a label set that represents relevant context-specific independencies (CSI), accounting for mechanism shifts induced by missingness. We define two causal effects within these systems: the \textit{full average treatment effect} (FATE), which reflects the effect in a hypothetical scenario \textit{had no data been missing}, and the \textit{natural average treatment effect} (NATE), which captures the effect  under the unaltered CSIs in the system. We propose recovery criteria for these queries and present doubly-robust estimators for a graphical model inspired by a real-world application. Simulations highlight key differences between these estimands and estimation methods. Findings from the application case suggest a small effect of ADHD treatment upon test achievement among Norwegian children, with a slight effect shift due to missing pre-tests scores.
\end{abstract}


\section{Introduction}\label{sec:intro}

In certain causality paradigms, the \textit{fundamental problem of causal inference} ---namely, the impossibility to simultaneously observe outcomes under both treatment and no treatment conditions--- has traditionally been regarded as a missing data problem \parencite{Neyman1923,rubin74, causalMissing}. Over the past decade, a growing perspective has taken shape in the opposite direction, recasting missing data problems as instances of causal inference. This viewpoint has led to an expanding body of research that combines methodologies from both areas \parencite{Mohan2013, Mohan2014, saadati2019adjustment, Bhattacharya2020, mgraphs, nabi2022causal}.

A central focus of this research has been to determine the conditions under which causal and counterfactual queries can be recovered in the presence of missing data, by combining graphical assumptions about the underlying causal mechanisms with information from potential auxiliary experiments \parencite{Mohan2013,Bhattacharya2020, mgraphs,NabiSemi}. Common queries include the \textit{average treatment effect} (ATE) of an exposure variable upon an outcome variable \parencite{Huber,ours}, the \textit{conditional average treatment effect} (CATE) and other functionals conditioned on evidence \parencite{tikka2021causal, kuzmanovic2023}, the interventional distribution \parencite{Mohan2013, Bhattacharya2020}, and the full data distribution \parencite{NabiFull}, among others.

Various techniques have been developed to recover and estimate causal effects under missing data and other forms of endogenous selection of unit information. These techniques include imputation methods \parencite{Rubin1976,RubinMIpaper,kyono2021miracle}, covariate adjustment \parencite{Correa_Tian_Bareinboim_2018, saadati2019adjustment, mathur2023}, \textit{inverse probability weighting} (IPW) or other ratios \parencite{HuberDW,Mohan2014}, and doubly-robust approaches \parencite{Wei2022,dw}. In certain settings, especially under a semiparametric model and missing outcomes, it has been shown that multiply-robust and efficient estimators for the ATE can incorporate elements from imputation, covariate adjustment, and weighting, thereby providing a unifying framework for estimation \parencite{Huber, ours}.

In methodological and applied research, special attention has been given to cases of missing outcome data, as these can arise from common phenomena such as attrition or loss to follow-up, which can threaten the validity of causal claims in both observational and experimental studies \parencite{hernan2004structural,attrition, biele2019bias}. The problems of recovering causal queries from missing covariate data \parencite{missConfounder, missingCov, Lewis2024} and missing exposure data \parencite{kuzmanovic2023,shi2024missingexposure} have also been studied, though aside from \textcite{saadati2019adjustment}, these works generally place less emphasis on graphical criteria. Furthermore, sound recovery algorithms and heuristics have been developed for general queries under multiple missingness and selection mechanisms \parencite{Bhattacharya2020, tikka2021causal}.

Several sets of sufficient conditions have been proposed for recovering causal effects from missing data. These typically combine: \textit{(i)} the existence of an admissible adjustment set that, along with relevant missingness indicators, can block all backdoor paths between exposure and outcome in an $m$-graph ---a graphical model encoding both causal relationships and missingness mechanisms; \textit{(ii)} specific $d$-separation conditions between the missingness indicators and either the exposure or the outcome; \textit{(iii)} a \textit{missing at random} (MAR) assumption for the variables affected by missingness; and \textit{(iv)} the requirement that missingness indicators are not causal descendants of the exposures transmitting the effect to the outcome, among other structural restrictions \parencite{saadati2019adjustment, mgraphs}. These conditions ensure that causal effect estimands, whether via covariate adjustment or IPW, remain valid despite the presence of missing data. Violations, by contrast, can induce selection bias and undermine causal conclusions for the target population \parencite{hernan2004structural}.

In most typical setups, if conditions for recovery are satisfied, the interpretation of the recovered effect is framed in an idealized scenario where no data are missing. For instance, a recovered ATE would be understood as the population-level expected difference in outcomes between treated and untreated scenarios, \textit{ceteris paribus}, and \textit{as if no data had been missing}  \parencite{nabi2022causal}. This interpretation allows the problem of inference under missing data to be viewed in a manner akin to a counterfactual scenario.

\paragraph{Motivation} Conventional $m$-graph-based approaches to missing data typically assume that missingness mechanisms affect only each other and do not have a direct causal impact on the substantive variables in the system \parencite{mgraphs, testability2, nabi2022causal}. In other words, the missingness status of a variable does not alter the causal mechanisms governing substantive variables. For the problem of missing covariate data, this assumption helps ensuring that the causal mechanisms of the exposure, the outcome, and their relationship remain invariant to the missingness indicators.

\begin{figure}[t]
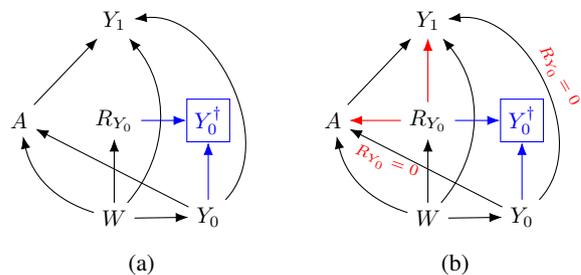

\centering
\begin{subfigure}{.5\linewidth}
  \centering
\tikz[scale=0.85, transform shape]{
    \node (a) {$A$};
    \node (r0) [right = of a, xshift = -2mm] {$R_{Y_0}$};
    \node (w) [below = of r0] {$W$};
    \node (yobs) [block, blue, right = of r0, xshift = -3mm] {$Y^\dagger_0$};
    \node (y0) [below = of yobs, yshift = 1mm] {$Y_0$};
    \node (y1) [above = of r0] {$Y_1$};

    \path (w) edge[style=directed] (y0);
    \path (w) edge[style=directed] (r0);
    \path (w) edge[style=directed, bend left=30] (a);
    \path (w) edge[style=directed, bend right=40] (y1);
    \path (y0) edge[style=directed, blue] (yobs);
    \path (r0) edge[style=directed, blue] (yobs);
    \path (a) edge[style=directed] (y1);
    \path (y0) edge[style=directed, bend right=70]  (y1);
    \path (y0) edge[style=directed]  (a);
}
    \caption{}
    \label{fig1b}
\end{subfigure}%
\begin{subfigure}{.5\linewidth}
  \centering
\tikz[scale=0.85, transform shape]{
    \node (a) {$A$};
    \node (r0) [right = of a, xshift = -2mm] {$R_{Y_0}$};
    \node (w) [below = of r0] {$W$};
    \node (yobs) [block, blue, right = of r0, xshift = -3mm] {$Y^\dagger_0$};
    \node (y0) [below = of yobs, yshift = 1mm] {$Y_0$};
    \node (y1) [above = of r0] {$Y_1$};

    \path (w) edge[style=directed] (y0);
    \path (w) edge[style=directed] (r0);
    \path (w) edge[style=directed, bend left=30] (a);
    \path (w) edge[style=directed, bend right=40] (y1);
    \path (y0) edge[style=directed, blue] (yobs);
    \path (r0) edge[style=directed, blue] (yobs);
    \path (r0) edge[style=directed, red] (y1);
     \path (r0) edge[style=directed, red] (a);
    \path (a) edge[style=directed] (y1);
    \path (y0) edge[style=directed, bend right=70] node[above, rotate=-60, red]{\scriptsize $R_{Y_0}=0$} (y1);
    \path (y0) edge[style=directed] node[above, xshift = -0.7cm, yshift= -0.15cm, rotate=-20, red]{\scriptsize $R_{Y_0}=0$} (a);
}
    \caption{}
    \label{fig1c}
\end{subfigure}
\caption{Graphical representations of systems with missing data on covariate $Y_0$ (a confounder of the casual relationship between $A$ and $Y_1$): (a) An $m$-graph including the proxy variable $Y^\dagger_0$, represented in a \textcolor{blue}{blue} box as a deterministic function of $R_{Y_0}$ and $Y_0$. (b) An $lm$-graph illustrating labeled CSIs, where $Y_0$ is an input for the mechanisms of $A$ and $Y_1$ when observed ($R_{Y_0} = 1$), but it is not when missing (\textcolor{red}{$R_{Y_0} = 0$}). Consequently, $R_{Y_0}$ becomes a causal parent of $A$ and $Y_1$.}
\label{fig1}
\end{figure}

However, in certain settings and applications, this constraint may be too restrictive or misaligned with the underlying data-generating process. Consider an observational study where schoolchildren take standardized tests at two consecutive time points, with corresponding scores $Y_0, Y_1 \in \mathbb{R}$. Between the two tests, some children receive stimulant medication, represented by the binary treatment indicator $A$. The goal is to estimate the ATE of stimulant medication on the second test score. Crucially, some students do not take the first test, indicated by the missingness indicator value $R_{Y_0} = 0$. If $Y_0$ is a key predictor of treatment assignment, it is plausible that units with missing test scores follow an alternative decision-making process regarding treatment, even after adjusting for other observed covariates. In other words, parents and clinicians of schoolchildren with missing first test scores may rely on different treatment decision rules than those who do observe $Y_0$. Furthermore, students who took the first test may be more familiar with the testing environment, potentially influencing their performance on the second test. This could result in systematic outcome differences between students who completed the first test and those who did not, even after accounting for other observed factors. In this setting, the missingness indicator directly influences both the exposure and the outcome, violating the conventional assumptions of $m$-graph models (see \cref{fig1b,fig1c}) and potentially inducing significant downstream shifts.

\paragraph{Prior work} Early methodological and applied work on missing covariate data highlighted the value of including interaction terms with missingness indicators to enhance inference robustness against potential mechanism shifts \parencite{greenland1995critical, jones1996indicator}. This is known as the \textit{missingness indicator method} (MIM), and extensive simulation studies have demonstrated that this approach is \textit{nearly valid} in most common applied scenarios \parencite{song2021, missingCov}. However, the MIM is not explicitly grounded in graphical criteria.

Recent work has explored various challenges at the intersection of missing data and mechanism shifts. For instance, \textcite{zhou2023domain} develop domain adaptation techniques for settings where missingness mechanisms differ across environments (e.g., between source and target populations), while the distributions of substantive variables remain stable. In the context of CATE generalization with missing exposure data, \textcite{kuzmanovic2023} study the impact of covariate shifts between treated and untreated groups, deriving generalization bounds to assess their influence. Addressing a similar generalization problem with missing covariates, \textcite{ColnetJosse,missingcovariates} propose imputation strategies and sensitivity analyses to manage distributional changes. While some of these approaches are inspired by graphical representations of the system, they typically do not employ graphical criteria directly in their solutions.

As part of a broader effort to extend $m$-graph-based models, \textcite{entangled} relax standard assumptions by introducing various forms of \textit{entanglement}, which capture classical unit interference as well as cases where one unit’s missingness mechanism can influence proxy covariates of other units. However, to the best of our knowledge, no previous work has directly addressed the problem of recovering causal effects in the presence of missingness-specific independencies and mechanism shifts that affect downstream variables; nor has prior work developed semiparametric theory-based estimators tailored to such settings.

\paragraph{Contributions} This paper addresses a limitation in current $m$-graph-based frameworks in causal inference with missing data, which do not allow for substantive descendants of the missingness indicators. As a result, $m$-graphs fail to capture missingness-induced mechanism shifts, which occur when missing data induce changes in downstream causal mechanisms. Empirical methods, such as the MIM, recommend the explicit modeling of interactions between missingness indicators and other covariates to make inference more robust to such shifts. However, these methods, despite being noted for their practical validity \parencite{song2021, missingCov}, lack a solid theoretical justification within the graphical framework. This gap is addressed in this paper by introducing $lm$-SCMs and $lm$-graphs, which expand on the existing $m$-graph models by incorporating a label set with relevant context-specific independencies (CSI). We provide a theoretical justification for these systems as resulting from latent soft interventions, and propose recovery criteria that integrate conditions from both $m$-graph and labeled graph ($l$-graph) frameworks. These contributions advance the field of causal inference with missing data, addressing a critical gap in the literature.


\section{Preliminaries}\label{sec:prelim}

\paragraph{Graph operations} Consider a \textit{directed acyclic graph} (DAG) $\mathcal{G}$ defined on nodes $\mathcal{V}$, and a subset $X \subseteq \mathcal{V}$. The notations $\pa(X;\mathcal{G})$, $\ch(X;\mathcal{G})$, $\an(X;\mathcal{G})$, and $\de(X;\mathcal{G})$ represent the parents, children, ancestors, and descendants of $X$ in $\mathcal{G}$, respectively. The mutilated graph $\mathcal{G}[\underline{X}]$ is obtained by removing all outgoing edges from the nodes in $X$, whereas $\mathcal{G}[\overline{X}]$ is formed by eliminating all incoming edges to $X$. 

\paragraph{Causal graphs} A \textit{structural causal model} (SCM) is a tuple $\mathcal{M} = (\mathcal{V}, \mathcal{U}, \mathcal{G}, \mathcal{F}, P_\mathcal{U})$. Here $\mathcal{V}$ represents a finite set of relevant variables; $\mathcal{U}$ is a finite set of noise variables; and $\mathcal{G}$ is a DAG over $\mathcal{V}$. The component $P_\mathcal{U}$ specifies a probability distribution for $\mathcal{U}$. The set $\mathcal{F} = \{f_V\}_{V \in \mathcal{V}}$ comprises a collection of measurable functions that describe the direct causal mechanisms: for each $V \in \mathcal{V}$, there exists an associated $U_V \in \mathcal{U}$ and a function $f_V : \supp \pa(V; \mathcal{G}) \times \supp U_V \to \supp V$ such that $V = f_V(\pa(V; \mathcal{G}), U_V)$ almost surely \parencite{PearlCausality}. Here, we denote with $\supp X$ the support of random variable $X$. In \textit{semi-Markovian} models, bidirectional arrows $V_1\leftrightarrow V_2$ indicate latent confounding  $V_1\!\leftarrow\! U\! \rightarrow\! V_2$, and $\mathcal{G}$ becomes an \textit{acyclic directed mixed graph} (ADMG) \parencite{evans2014markovian}.

Let $A$ and $Y$ denote the \textit{exposure} and the \textit{outcome}, respectively. The \textit{unit-level counterfactual} or \textit{potential outcome}, $Y^a(u)$, represents the value that $Y$ would take if an intervention were performed by setting $A$ to a fixed value $a \in \supp A$ for an individual characterized by $\mathcal{U} = u$ in the SCM $\mathcal{M}$. This intervention propagates through the system, updating descendant variables according to the causal mechanisms $\mathcal{F}$ applied in a topological order dictated by $\mathcal{G}$. Potential outcomes satisfy the consistency axiom: if $A(u) = a$ and $Y(u) = y$, then $Y^a(u) = y$ \parencite{robinsConsistency, PearlCausality}. The corresponding population-level distribution, known as the \textit{interventional distribution}, is expressed as $p(y \mid \doo(A = a)) = \int \I\left\{u \in \mathcal{U}^a[y]\right\} \, \dd P(u)$, where\\  $\mathcal{U}^a[y] = \{u \in \supp \mathcal{U} : Y^a(u) = y\}$ is the pre-image of $y \in \supp Y$ under $Y^a(\cdot)$ \parencite{BareinboimHierarchy2022}.

The \textit{average treatment effect} (ATE), $\psi$, is one of the most commonly studied \textit{causal effects} or \textit{queries}. For a binary point-exposure $A \in \{0,1\}$ and a continuous outcome $Y \in\mathcal{Y}\subseteq \mathbb{R}$, the ATE is defined as:
\begin{equation}\label{eq:ATE}
    \psi := \Delta_a \E\left[Y \mid \doo(A = a)\right],
\end{equation}

where $\Delta_a$ denotes the difference operator relative to $a$.

A sound and complete set of three rules known as the \textit{$do$-calculus} aids in the identification of interventional distributions by leveraging the conditional independencies implied by directional separation (\textit{$d$-separation}) statements embedded in $\mathcal{G}$ and its mutilations  \parencite{Pearl1995,PearlDo}. 

\paragraph{Soft interventions}  
\textit{Soft\,/\,stochastic interventions} extend \textit{do}-actions by enabling surgical modifications of mechanisms within an SCM $\mathcal{M}$ with causal graph $\mathcal{G}$. Instead of fixing variables to specific values, they replace original mechanisms with alternative functions and inputs. For a given $X \in \mathcal{V}$, a soft intervention $\sigma_X = (D, g, \widetilde{U})$ consists of new parents from its original nondescendants, $D \subseteq  \operatorname{nde}(X; \mathcal{G})$, a measurable function $g$, and an independent auxiliary noise $\widetilde{U}$ (which may be empty). This intervention induces a new SCM  $\mathcal{M}^{\sigma_X}$, where the original assignment $X  \!\leftarrow\! f_X(\pa(X;\mathcal{G}), U_X)$ is replaced with $X  \!\leftarrow\! g(D, U_X, \widetilde{U})$ almost surely. The corresponding \textit{operated graph} $\mathcal{G}^{\sigma_X}$ is obtained by removing edges $\pa(X;\mathcal{G}) \!\to\! X$, adding edges $D \!\to\! X$, and introducing a \textit{regime arrow} $\sigma_X \!\to\! X$. For foundational work on this topic, we refer the reader to \textcite{CorreaIjcai, Correa_Bareinboim_2020, NeuripsCorrea}.

\paragraph{Missingness graphs} Let $\mathcal{M}$ be an SCM, and  $\mathcal{V}_o\cup\mathcal{V}_m\cup\mathcal{R}$ be a partition of variables $\mathcal{V}$ in $\mathcal{M}$, where $\mathcal{V}_o$ contains fully observed variables, $\mathcal{V}_m$ contains variables affected by missing data, and $\mathcal{R}=\{R_V : V\in\mathcal{V}_m\}$ collects the missingness indicators. That is, if $V$ is observed for an individual unit, then $R_V=1$ for the same unit; and, if it is missing, then $R_V=0$. The causal graph $\mathcal{G}$ in $\mathcal{M}$ is called a missingness graph, or \textit{$m$-graph}, as it involves the indicators $\mathcal{R}$ in addition to the \textit{substantive variables} $\mathcal{V}_o\cup\mathcal{V}_m$. It is typically assumed that $\mathcal{R}$ has no substantive descendants in $\mathcal{G}$ \parencite{Mohan2014, mgraphs, testability2}. Let us denote with $V^\dagger$ the \textit{proxy} for $V\in\mathcal{V}_m$, such that $V^\dagger=V$ almost surely when $R_V=1$, and it takes the empty value $V^\dagger=\oldempty$ otherwise. The distribution $P_\mathcal{V}$ is known as the \textit{full data distribution}, while $P_{\mathcal{V}^\dagger}$, with $\mathcal{V}^\dagger=\mathcal{V}_o\cup\mathcal{V}_m^\dagger\cup\mathcal{R}$, is the \textit{observed data distribution}. In other formulations, $V$, $V^\dagger$ and $\oldempty$ are denoted respectively with $V^1$, $V$ and $?/\texttt{NA}$ \parencite{nabi2022causal}. The explicit $m$-graph can incorporate the proxy variables $V^\dagger$, highlighting their deterministic nature given their two parents $R_V$ and $V$.

A causal query $\Psi$ acting on SCMs, such as the interventional distribution or the ATE,  is \textit{(nonparametrically)} \textit{recoverable} from incomplete observational data if it outputs the same value $\psi$ for all models within the class of SCMs that share the same $m$-graph $\mathcal{G}$ and the same positive observed data distribution $\mathfrak{M}(\mathcal{G},P)$. That is, $\forall\mathcal{M}\in\mathfrak{M}(\mathcal{G},P),\, \Psi[\mathcal{M}]=\psi$. In other words, the query is uniquely computable as a functional of the graph and the observed data, yielding a recovered statistical \textit{estimand} \parencite{Bareinboim_Tian_Pearl_2014,mgraphs}. Graphical criteria can enable the recovery of causal queries even in complicated settings beyond the conventional MAR model \parencite{Mohan2013,NabiSemi}. Several sufficient criteria for recovery of causal effects have been proposed, some based on sequential factorizations and covariate adjustment \parencite{Mohan2014,Correa_Tian_Bareinboim_2018,saadati2019adjustment,ours}, and others based on ratio factorizations and IPW \parencite{horvitz1952generalization, HuberDW, Mohan2014}. More encompassing algorithmic approaches for assessing recoverability include the identification algorithm introduced by \textcite{Bhattacharya2020} and the search heuristic by \textcite{tikka2021causal}. 


\paragraph{Labeled graphs} A \textit{context-specific independence} (CSI) is a form of local statement that extends the concept of conditional independence by holding only in certain specific contexts \parencite{BoutilierCSI}. For mutually exclusive sets of variables $Z,X,B,C$, the expression $ Z \indep X \mid B, C = c $ represents a CSI, indicating that $Z$ and $X$ are conditionally independent given $B$ and the specific context $C = c$. 


A \textit{labeled graph}, $l$-graph, or LDAG, is a tuple $(\mathcal{G}, \mathcal{L})$ consisting of a DAG $\mathcal{G}$ and a \textit{label set} $\mathcal{L} = \bigcup_{C \in \mathscr{C}} \bigcup_{c \in \nu(C)} \mathcal{L}_C(c)$. Here, $\mathscr{C} \subset \mathcal{V}$ is a set of context variables,  $\nu(C) \subseteq \supp C$ is a set of context values for $C\in \mathscr{C}$, and $\mathcal{L}_C(c)$ contains a collection of edges $(X,Z)\in\operatorname{edg}(\mathcal{G})$ for which the CSI $Z \indep X \mid \pa(Z; \mathcal{G}) \setminus (X \cup C), C = c$ holds \parencite{Pensar2015, EhsanCSI}. If $(X, Z) \in \mathcal{L}_C(c)$, the label ``$C = c$'' can be graphically depicted on the arrow $X \rightarrow Z$.

An $l$-graph is called \textit{regular-maximal} if \textit{(i)} for all $(X,Z)\in\bigcup_{c \in \nu(C)} \mathcal{L}_C(c)$, one has that $\pa(Z;\mathcal{G})\cap C\neq\emptyset$, and \textit{(ii)} it is not possible to add an additional edge $(X,Z)\in\operatorname{edg}(\mathcal{G})$ to $\mathcal{L}$ without inducing a new CSI  \parencite{EhsanCSI}.

It is known that the \textit{do}-calculus can fail to identify causal effects in nontrivial $l$-graphs that are indeed identifiable. \textcite{TikkaNeurips} demonstrated that identifying causal effects from $l$-graphs is NP-hard and proposed a sound CSI-calculus along with a heuristic approach to obtain identifying formulas. Building on this, \textcite{EhsanCSI} showed that when all context variables are \textit{control} variables (i.e. having no parents in $\mathcal{G}$) a causal effect is identifiable from a regular-maximal $l$-graph if it can be identified from a series of causal graphs, each representing a different context world, where each identification problem can be assessed using existing sound and complete procedures.


\section{Labeled missingness graphs}\label{sec:lmgraphs}

To formally define the causal effects and their recovery conditions under missingness-specific independencies, we first introduce $lm$-SCMs and $lm$-graphs. For notational convenience, we omit curly brackets when referring to singleton sets, allowing us to write $X$ for $\{X\}$.

\begin{definition}[$lm$-SCM]\label{def:lmscm}
Let $\mathcal{M}$ be a semi-Markovian SCM with $m$-graph structure $\mathcal{G}^*$. Let the set of variables affected by missingness in $\mathcal{M}$ be partitioned as $\mathcal{V}_m = \mathcal{V}_{\rm{no}} \cup \mathcal{V}_{\rm{sh}}$, where $\mathcal{V}_{\rm{no}}$ are variables whose missingness does not induce mechanism shifts, and $\mathcal{V}_{\rm{sh}}$ are those whose missingness may induce mechanism shifts. Let 
$\mathcal{R} = \mathcal{R}_{\rm{no}} \cup \mathcal{R}_{\rm{sh}}$ be the corresponding partition for the missingness indicators.

For each variable $X \in \mathcal{V}_{\rm{sh}}$:
\begin{itemize}[leftmargin=*]
    \item Let $\mathcal{S}_X \subseteq \ch(X;\mathcal{G}^*) \setminus \an(R_X;\mathcal{G}^*)$ be a nonempty set of selected children of $X$ that are not ancestors of $R_X$, representing the \textbf{shifted children} of $X$.
\end{itemize}

For any variable $Z \in \mathcal{V}$:
\begin{itemize}[leftmargin=*]
    \item Define the set of \textbf{contextual parents} of $Z$ as $\mathcal{T}_Z := \{X \in \mathcal{V}_{\rm{sh}} : Z \in \mathcal{S}_X\}$.
    \item For $Z$ with $\mathcal{T}_Z \neq \emptyset$, let $\sigma_Z$ be a \emph{soft intervention} that modifies the natural assignment $Z\leftarrow f_Z(\pa(Z;{\mathcal{G}^*}), U_Z)$ governing $Z$ in the original SCM $\mathcal{M}$ to be:
\end{itemize}
\begin{align*} 
         Z\leftarrow & \left[\prod_{X\in \mathcal{T}_Z}R_X\right] f_Z(\pa(Z;{\mathcal{G}^*}), U_Z) + \\ \notag
    & \sum_{\substack{T\subseteq \mathcal{T}_Z\\ T\neq\emptyset}}
    \left[\prod_{X\in T}(1-R_X)\right]\, 
    g_{Z,T}(\pa(Z;{\mathcal{G}^*})\setminus T, U_Z),
\end{align*}

where each $g_{Z,T}$ is a measurable function that defines a shifted mechanism when data from a subset $T$ of contextual parents are missing.

The SCM $\mathcal{M}^\sigma$ induced by joint soft interventions $\sigma=\{\sigma_Z : \mathcal{T}_Z \neq \emptyset\}$ is referred to as a labeled missingness SCM.
\end{definition}


An $lm$-SCM can represent systems such as the one described in the motivation, where some variables $Z$, directly influenced by variables affected by missingness $\mathcal{T}_Z$, rely on an alternative mechanism that excludes some contextual parent variables when their values are inaccessible. 

\begin{definition}[$lm$-graph]\label{def:lmgraph} A \textit{labeled missingness graph}, or $lm$-graph, representing an $lm$-SCM $\mathcal{M}^\sigma$, is a tuple $(\mathcal{G}, \mathcal{L})$, where these components can be constructed as:
\begin{enumerate}[label=(\roman*)]
    \item $\mathcal{G}$ starts with the same structure as the underlying $m$-graph $\mathcal{G}^*$, including bidirectional arrows, and, for all $X\in\mathcal{V}_{\rm{sh}}, Z\in \mathcal{S}_X$, it includes the arrows $R_X\rightarrow Z$,
    \item $\mathcal{L}=\bigcup_{R\in\mathcal{R}_{\rm{sh}} } \mathcal{L}_{R}$, where $\mathcal{L}_{R_X}=\{(X,Z) : Z\in \mathcal{S}_X \wedge X\not\leftrightarrow Z \text{ in } \mathcal{G}\}$,
    \item For all $(X,Z)\in\mathcal{L}$, the label ``$R_X= 0$'' can be graphically depicted on the arrow $X \rightarrow Z$.
    \end{enumerate}
\end{definition}

\begin{figure}[t]
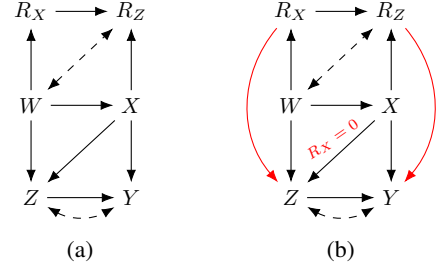

\centering
\begin{subfigure}{.2\textwidth}
  \centering
\tikz[scale=0.825, transform shape]{
    \node (w) {$W$};
    \node (x) [right = of w] {$X$};
    \node (z) [below = of w] {$Z$};
    \node (y) [below = of x] {$Y$};

    \node (rx) [above = of w] {$R_X$};
    \node (rz) [above = of x] {$R_Z$};

    \path (w) edge[style=directed] (x);
    \path (w) edge[style=directed] (z);
    \path (w) edge[style=directed] (rx);
    \path (x) edge[style=directed] (rz);
    \path (rx) edge[style=directed] (rz);
    \path (x) edge[style=directed] (y);
    \path (x) edge[style=directed] (z);
    \path (z) edge[style=directed] (y);
    \path (w) edge[style=bidirected] (rz);

    \path (z) edge[style=bidirected, bend right=30] (y);
}
    \caption{}
    \label{fig2a}
\end{subfigure}%
\begin{subfigure}{.2\textwidth}
  \centering
\tikz[scale=0.825, transform shape]{
    \node (w) {$W$};
    \node (x) [right = of w] {$X$};
    \node (z) [below = of w] {$Z$};
    \node (y) [below = of x] {$Y$};

    \node (rx) [above = of w] {$R_X$};
    \node (rz) [above = of x] {$R_Z$};

    \path (w) edge[style=directed] (x);
    \path (w) edge[style=directed] (z);
    \path (w) edge[style=directed] (rx);
    \path (x) edge[style=directed] (rz);
    \path (rx) edge[style=directed] (rz);
    \path (x) edge[style=directed] (y);
    \path (x) edge[style=directed] node[above, xshift = -0.0cm, yshift= 0.01cm, rotate=35, red]{\scriptsize $R_X=0$} (z);
    \path (z) edge[style=directed] (y);
    \path (w) edge[style=bidirected] (rz);

    \path (z) edge[style=bidirected, bend right=30] (y);
    \path (rx) edge[style=directed, red, bend right=40] (z);
    \path (rz) edge[style=directed, red, bend left=40] (y);
}
    \caption{}
    \label{fig2b}
\end{subfigure}%
\caption{Construction of an $lm$-graph with $\mathcal{S}_X = Z$ and $\mathcal{S}_Z = Y$: (a)  Underlying $m$-graph $\mathcal{G}^*$; (b)  $lm$-graph $\mathcal{G}$ introducing the arrows $R_X \to Z$ and $R_Z \to Y$. Since there is no latent confounding between $Z$ and $X$ in $\mathcal{G}$, the edge $X \to Z$ is labeled with ``$R_X = 0$''. In contrast, no label is assigned to the edge $Z \to Y$ because latent confounding is present.}
\label{fig2}
\end{figure}

The addition of new arrows to causal graphs carries the risk of creating cycles, which would prevent the system from being represented by a DAG or an ADMG and lead to problematic interpretations of circular causality. That issue is avoided here by requiring that, for any variable $X$, its children that are ancestors of $R_X$ are never shifted by $R_X$. This restriction effectively prevents feedback loops. \Cref{fig2} illustrates the construction of an $lm$-graph from the underlying $m$-graph and sets of shifted nodes.

An $lm$-graph combines the graphical expressiveness of both $m$-graphs and $l$-graphs, offering a unified framework to represent the problem at hand. Notably, when $\mathcal{V}_{\rm{sh}} = \emptyset$, the $lm$-graph simplifies to an $m$-graph. Additionally, the following two remarks characterize them as $l$-graphs.

\begin{proposition}\label{prop:1} Let $(\mathcal{G},\mathcal{L})$ be an $lm$-graph representing a semi-Markovian $lm$-SCM, with $(X,Z)\in\mathcal{L}$, and let $\mathcal{G}_{\setminus (X,Z)}$ be the graph with the edge $X\rightarrow Z$ removed from $\mathcal{G}$. If $Z \indep_d X\mid W,R_X$ in $\mathcal{G}_{\setminus (X,Z)}$ then:
\begin{equation*}
Z \indep X \mid W, R_X = 0.
\end{equation*}
\end{proposition}

\textit{Proof}: Let $(X, Z) \in \mathcal{L}$, and suppose $X$ and $Z$ are $d$-separated in $\mathcal{G}_{\setminus (X,Z)}$ when conditioning on $R_X$ and a disjoint set of variables $W$. Then, this $d$-separation statement constitutes a CSI-separation. By the soundness of CSI-separation \parencite{koller2009probabilistic, logicalCSI}, it follows that $Z \indep X \mid W, R_X = 0$. Since the label set $\mathcal{L}$ encodes such CSI statements, $lm$-graphs are $l$-graphs, in which conditional independence under specific contexts can be assessed via $d$-separation after removing the labeled edges corresponding to those contexts. \hfill$\square$

\begin{proposition}\label{prop:2}  $lm$-graphs are regular-maximal $l$-graphs.
\end{proposition}

\textit{Proof}: If $(X, Z) \in \mathcal{L}$, then the edge $R_X \to Z$ is present in $\mathcal{G}$, ensuring regularity as $\pa(Z; \mathcal{G}) \cap R_X \neq \emptyset$. Since each variable in $\mathcal{R}_{\rm{sh}}$ is binary, it follows from Corollary 1 in \textcite{EhsanCSI} that $lm$-graphs are maximal. That is because it is impossible to add $(X, Z)$ to $\mathcal{L}$ more than once from a different context where $R_X =r\neq 0$. \hfill$\square$

Consider the graph in \cref{fig1c}. It is an example of an $lm$-graph with $\mathcal{R}_{\rm{sh}} = R_{Y_0}$, $\mathcal{S}_{Y_0} = \{A, Y_1\}$, and $\mathcal{L} = \{(Y_0, A), (Y_0, Y_1)\}$. Building on the motivating case from the \nameref{sec:intro}, it represents an $lm$-SCM in which, when initial test scores $Y_0$ are missing ($R_{Y_0} = 0$), parents and clinicians of schoolchildren rely on a different  rules for prescribing ADHD medication ($A$) compared to those who observe and leverage the unit's information on $Y_0$. The assignment mechanism for $A$ is then modified by a latent soft intervention \parencite{Correa_Bareinboim_2020} to be:
\begin{equation*}
    A\leftarrow R_{Y_0}\cdot f_A(W,Y_0, U_A) + (1-R_{Y_0})\cdot g_A(W, U_A),
\end{equation*}
where $g_A$ is a shift function. Clearly, when $R_{Y_0} = 0$, the causal mechanisms for $A$ no longer uses $Y_0$ as an input. Since the underlying graph contains no latent confounding $A\leftrightarrow Y_0$, this leads to the CSI $A \indep Y_0 \mid W, R_{Y_0} = 0$. 

Furthermore, schoolchildren who took the test at the initial time point may have gained more experience in test-taking, potentially affecting their performance even after accounting for all other relevant factors. Thus, a similar argument leads to the other CSI in \cref{fig1c}, $Y_1 \indep Y_0 \mid W, A, R_{Y_0} = 0$.

We now introduce causal queries defined within $lm$-SCMs.

\section{Causal Effects on $lm$-graphs}\label{sec:causal}

In this section, we examine the problem of defining, interpreting, and recovering causal effects from partially observed data generated by an $lm$-SCM $\mathcal{M}^\sigma$. We work under the assumption that the outcome variable $Y$ has no causal descendants in the system.

Within the $lm$-SCM framework, and for a binary point exposure $A \in \{0,1\}$ and a continuous outcome $Y  \in \mathbb{R}$, different causal queries analogous to the ATE, as defined in \cref{eq:ATE}, can be formulated. To clarify these distinctions, we introduce two specific versions: the FATE and the NATE.

\begin{definition}[FATE]\label{def:rate} The \textit{full average treatment effect} (FATE) is the population-level expected difference in outcomes between treated and untreated scenarios, \textit{ceteris paribus}, had no data been missing, under $\mathcal{M}^\sigma$:
\begin{equation*}
    \phi := \Delta_a \E\left[Y \mid \doo(A = a,\mathcal{R}=\boldsymbol{1})  \right],
\end{equation*}

where $\mathcal{R}= \boldsymbol{1}$ is a short notation for $R=1,\forall R\in\mathcal{R}$.
\end{definition}

The FATE corresponds to a hypothetical ATE in a scenario with full information observability and no mechanism shifts. The following proposition outline insights for its recovery.



\begin{proposition}\label{prop:X} Let $\mathcal{R}_\Phi\subseteq \mathcal{R}_{\rm{sh}}$ be the indicators connected to $Y$ via directed (causal) paths that do not intersect $\mathcal{R}_{\rm{sh}}\cup A$, as intermediate nodes, in the $lm$-graph $(\mathcal{G},\mathcal{L})$, that is:
\begin{equation*}
    \mathcal{R}_\Phi := \mathcal{R}_{\rm{sh}} \cap \an(Y; \mathcal{G}[\overline{\mathcal{R}_{\rm{sh}}\cup A}]).
\end{equation*}

Then, for binary exposure $A$, the FATE is equivalent to:
\begin{equation}\label{eq:equality}
     \phi = \Delta_a \E\left[Y \mid \doo(A = a,\mathcal{R}_\Phi=\boldsymbol{1})  \right],
\end{equation}

and recoverable from $(\mathcal{G},\mathcal{L})$ if $P(Y \mid \doo(A,\mathcal{R}_\Phi))$ is recoverable from $\mathcal{G}$ treated as an unlabelled $m$-graph with substantive descendants of $\mathcal{R}$.
\end{proposition}


\textit{Proof}: The intervention $\doo(\mathcal{R}_{\rm{sh}}=\boldsymbol{1})$ makes the labels in $\mathcal{L}$ idle, so the analysis can be conducted using only $\mathcal{G}$. By Rule 3 of \textit{do}-calculus one has that $P(Y \mid \doo(A,\mathcal{R}))=$ $P(Y \mid \doo(A,\mathcal{R}_{\rm{sh}})) $ whenever $Y \indep_d \mathcal{R}_{\rm{no}} \mid A, \mathcal{R}_{\rm{sh}}$ in $\mathcal{G}[\overline{A,\mathcal{R}}]$ \parencite{PearlCausality}. Nodes in $\mathcal{R}_{\rm{no}}$ have no parents in $\mathcal{G}[\overline{A,\mathcal{R}}]$ and, since their only possible children belong to $\mathcal{R}$, they have no descendants either. As a result, they are isolated from $Y$ in such graph. Plus, all directed causal paths from nodes $\mathcal{R}_{\rm{sh}}\setminus\mathcal{R}_\Phi$ to $Y$ are intersected by $\mathcal{R}_\Phi$ or $A$, so by Rule 3 $P(Y \mid \doo(A,\mathcal{R}_{\rm{sh}}))= P(Y \mid \doo(A,\mathcal{R}_\Phi))$. \hfill$\square$

The set $\mathcal{R}_\Phi$ consists precisely of those missingness indicators that have a causal path to $Y$ under intervention on both $A$ and $\mathcal{R}$; these indicators trigger mechanism shifts and can transmit both their own effects and the effect of the exposure to the outcome $Y$. This implies that recovering the FATE from observed data, $(\mathcal{V}_o, \mathcal{V}_m^\dagger, \mathcal{R}) \sim P^\sigma$, hinges on the recoverability status of $P(Y \mid \doo(A, \mathcal{R}_\Phi)) $ from $\mathcal{G}$ as an $m$-graph (with substantive descendants of $\mathcal{R}$). 

Various available methods can be employed to assess this task. For instance, in the scenario depicted in \cref{fig1c}, the FATE can be recovered through sequential factorizations \textcite{mgraphs}, leading to the expressions:  
\begin{align}\label{eq:rateeq1}
   &  \phi = \E_W\E_{Y_0\mid W,R_{Y_0}=1}\Delta_a  Q_1(W,Y_0,a), \text{ with}\\ \label{eq:rateeq2}
   &  Q_1(W,Y_0,A) = \E \left[Y\mid W,Y_0,A,R_{Y_0}=1 \right].
\end{align}

This result follows from verifying that $\mathcal{R}_\Phi=R_{Y_0}$, and that $P(Y_1 \mid \doo(A, R_{Y_0}))$ can be expressed as:
\begin{equation*}
    \int \dd P(W) \, \dd P(Y_0 \mid W, R_{Y_0}) \, P(Y_1 \mid W, Y_0, A, R_{Y_0}),
\end{equation*}

as the following sequence of graphical conditions hold:
\begin{enumerate}[label=(\roman*)]
    \item $ \{W, Y_0\} \cap \de(A\cup R_{Y_0}; \mathcal{G})=\emptyset $,
    \item $ Y_1 \indep_d A, R_{Y_0} \mid W, Y_0 $ in $\mathcal{G}[\underline{A,R_{Y_0}}]$, and
    \item $ Y_0 \indep_d R_{Y_0} \mid W $ in $\mathcal{G}$.
\end{enumerate}

\begin{figure}[t]
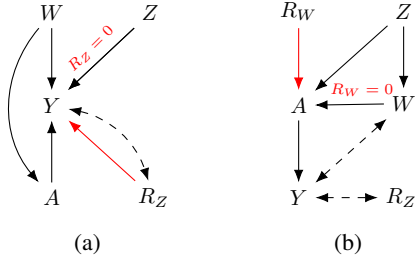

\centering
\begin{subfigure}{.2\textwidth}
  \centering
\tikz[scale=0.825, transform shape]{
    \node (y) {$Y$};
    \node (a) [below = of y] {$A$};
    \node (r) [right = of a] {$R_Z$};
    \node (w) [above = of y] {$W$};
    \node (z) [right = of w] {$Z$};
    
    \path (w) edge[style=directed] (y);
    \path (z) edge[style=directed] (y);
    \path (a) edge[style=directed] (y);
    \path (y) edge[style=bidirected, bend left=30] (r);
    \path (w) edge[style=directed, bend right=40] (a);
    \path (z) edge[style=directed] node[above, xshift = -0.0cm, yshift= 0.01cm, rotate=45, red]{\scriptsize $R_Z=0$} (y);
    \path (r) edge[style=directed, red] (y);

}
    \caption{}
    \label{fig3b}
\end{subfigure}%
\begin{subfigure}{.2\textwidth}
  \centering
\tikz[scale=0.825, transform shape]{
    \node (a) {$A$};
    \node (y) [below = of a] {$Y$};
    \node (r) [above = of a] {$R_W$};
    \node (w) [right = of r] {$Z$};
    \node (u2) [below = of w] {$W$};
    \node (u1) [right = of y] {$R_Z$};

    \path (r) edge[style=directed, red] (a);
    \path (a) edge[style=directed] (y);

    \path (w) edge[style=directed] (a);
    \path (w) edge[style=directed] (u2);
    \path (u2) edge[style=directed] node[above, xshift = 0.2cm, yshift= +0.01cm, red, rotate=0]{\scriptsize $R_W=0$} (a);
    \path (u2) edge[style=bidirected] (y);
    \path (u1) edge[style=bidirected] (y);

}
    \caption{}
    \label{fig3c}
\end{subfigure}%
\caption{Contrasting recovery results from $lm$-graphs versus underlying $m$-graphs (where the \textcolor{red}{red} edges and labels are absent): (a) In the $m$-graph, $P(Y \mid \doo(A))$ is recoverable because $Z$ is not needed for adjustment, allowing recovery via identification $\int \dd P(W) P(Y \mid W, A)$. In contrast, $P(Y \mid \doo(A, R_Z))$ is not recoverable from the $lm$-graph due to latent confounding $Y \leftrightarrow R_Z$; (b) Here, $P(Y \mid \doo(A))$ is not recoverable from the $m$-graph because $Z$ cannot be used to block the backdoor path $A \leftarrow Z \rightarrow W \leftrightarrow Y$ as $R_Z$ is not $d$-separable from $Y$. However, in the $lm$-graph, it is recoverable by applying Rule 1 of \textit{do}-calculus: $P(Y \mid \doo(A)) = P(Y \mid \doo(A), R_W)$, and given no confounding when $R_W = 0$, this simplifies to $P(Y \mid A, R_W = 0)$.}
\label{fig3}
\end{figure}

Although the statistical estimand for the FATE $\phi$ in \cref{eq:rateeq1,eq:rateeq2} matches the expression obtained by recovering the ATE from the underlying $m$-graph, these do not always coincide. In some cases, the ATE may be recoverable from the underlying $m$-graph, while the FATE is not from the $lm$-graph, as illustrated in \cref{fig3b}.

For the FATE, a hypothetical intervention makes all variables observed by each unit, thereby avoiding any mechanism shifts that arise from incomplete information. In contrast, a different causal query is the \textit{natural} ATE (NATE\footnote{  The term “NATE” has also been used in recent literature to refer to the \textit{nudge} ATE: the average causal effect among the subgroup of units whose treatment can be manipulated by an instrumental variable \parencite{tchetgen2024nudge}.} ). When the system is faithfully described by an $lm$-graph and units adapt their behavior based on which inputs are missing, the NATE captures the causal effect while accounting for such missingness-driven shifts in the causal mechanism.


\begin{definition}[NATE]\label{def:pate} The \textit{natural average treatment effect} (NATE) is the population-level expected difference in outcomes between treated and untreated scenarios, \textit{ceteris paribus}, under $\mathcal{M}^\sigma$:
\begin{equation*}
    \theta := \Delta_a \E\left[Y \mid \doo(A = a)  \right].
\end{equation*}
\end{definition}


Clearly, the NATE is recoverable if $P(Y\mid \doo(A))$ is recoverable from the $lm$-graph. Since $lm$-graphs are regular-maximal, if all variables in $\mathcal{R}_{\mathrm{sh}}$ were exogenous, recovery could in principle be evaluated using a procedure akin to the one proposed by \textcite{EhsanCSI}. However, because the recovery problem is more complex than the identification problem, such procedure would provide only a sufficient, but not necessary, condition for recovery under missingness-specific independencies. For example, \cref{fig3c} illustrates a case in which $R_W$ is exogenous and $P(Y \mid \doo(A))$ is recoverable, yet recovery cannot be achieved through a context-by-context sequence. Instead, it requires applying the rules of $do$-calculus prior to selecting a representative context. Notably, in this example, $P(Y \mid \doo(A))$ is not recoverable from the underlying $m$-graph alone, but becomes recoverable from the $lm$-graph.

In more general settings, recovering the NATE from an $lm$-graph remains a challenging task. A comprehensive solution would likely require integrating the CSI-calculus introduced by \textcite{TikkaNeurips} with the algorithm for causal effect recovery under missing data proposed by \textcite{Bhattacharya2020}. However, such combination is not guaranteed to be complete, implying that there may exist recoverable cases that fall outside their scope. Moreover, \textcite{TikkaNeurips} demonstrate that the general identification problem becomes NP-hard when CSI constraints are present.

Recovery of the NATE depends on the recovery of the distribution $P(Y \mid \doo(A))$ from the $lm$-graph $(\mathcal{G}, \mathcal{L})$, where CSIs are non-idle. We now present a sufficient condition for the recovery of this interventional distribution.

\begin{proposition}\label{prop:4} 
Let $K, H \subseteq \mathcal{V}_{\mathrm{sh}} \setminus (A \cup Y)$ be disjoint sets of variables whose missingness induce shifts, with $K = \{K_j\}_{j=1}^\kappa$ indexed. Let $R_K$ and $R_H$ denote their missingness indicators. For every missingness pattern $r \in \operatorname{supp} R_K \subseteq \{0,1\}^\kappa$, let $L_r \subseteq \mathcal{V} \setminus (A \cup Y \cup H)$ be a set such that:
\begin{enumerate}[label=(\roman*)]
    \item $L_r \cap \bigcup_{j: r_j = 0}^\kappa K_j = \emptyset$ and 
    $\bigcup_{j: r_j = 1}^\kappa K_j \subseteq L_r$, \\
    (i.e., $L_r$ excludes variables $K_j$ with $r_j = 0$, and includes all $K_j$ with $r_j = 1$).
\end{enumerate}

Define $M_r := \mathcal{V}_m \cap \left( A \cup Y \cup L_r \setminus K \right)$. Let $\mathcal{G}_r$ be the graph obtained by removing from $\mathcal{G}$ all arrows $\bigcup_{j : r_j = 0} \mathcal{L}_{R_{K_j}}$ in the label set $\mathcal{L}$, and $\mathcal{H}_{r}$ be the graph that removes the edges $\mathcal{L}_{R_H}$ from $\mathcal{G}_r$. Let the following conditions hold for every missingness pattern $r \in \operatorname{supp} R_K$:
\begin{enumerate}[label=(\roman*)]\setcounter{enumi}{1}
    
    \item $(R_K \cup L_r \cup R_{M_r} \cup R_H) \cap \operatorname{de}(A;\mathcal{G}_r) = \emptyset$,
    
    \item $Y \indep_d R_{M_r} \mid A, R_K$ in $\mathcal{G}_r[\overline{A}]$, 
    
    \item $Y \indep_d R_H \mid A, L_r, R_K, R_{M_r}$ in $\mathcal{G}_r[\overline{A}]$, 
    
    \item $Y \indep_d A \mid L_r, R_K, R_{M_r}, R_H$ in $\mathcal{H}_r[\underline{A}]$.
\end{enumerate}

If $K,H$ and $\{L_r\}_r$ exist and fulfill conditions above then:
\begin{align}\label{eq:nate1}
    & P(Y\mid\doo(A=a)) =\!\!\! \sum_{r\in\supp R_K}\mathbb{P}(R_k=r)\,\Theta_r(a),\\ \label{eq:nate2}
    & \Theta_r(a) = \int \dd P(L_r\mid R_K=r,R_{M_r}=1)\\ \notag
    &\qquad\qquad P(Y\mid A=a,L_r,R_k=r,R_{M_r}=1, R_H=0)
\end{align}

and thus the NATE, $\theta$, is given by:
\begin{align}\label{eq:nate}
    & \theta = \sum_{r\in\supp R_K}\mathbb{P}(R_k=r)\,\Delta_a\vartheta_r(a), \text{ where}\\
    & \notag \vartheta_r(a) = \int \dd P(L_r\mid R_K=r,R_{M_r}=1),\\ \notag
    &\qquad\qquad \E[Y\mid A=a,L_r,R_k=r,R_{M_r}=1, R_H=0].
\end{align}
\end{proposition}
 
\textit{Proof}: For any set $K\subseteq\mathcal{V}_{\rm{sh}}\setminus (A\cup Y)$, one can express\\ $P(Y \mid \doo(A))\!=\!\sum_{R_K}\! \mathbb{P}(R_K \mid \doo(A)) \, P(Y \mid \doo(A), {R}_K)$. Given a pattern $R_K = r$, terms conditioned in this context can be evaluated within the corresponding graph $\mathcal{G}_r$, obtained by removing the edges associated with the labels ``$R_{K_j} = r_j = 0$''. By assumption \textit{(ii)}, the first factor simplifies as $\mathbb{P}({R}_K)$. By  \textit{(iii)} and Rule 1 of \textit{do}-calculus, the second factor is  $P(Y \mid \doo(A), {R}_K)= \int \dd P(L_r\!\mid\!\doo(A),R_K,R_{M_r}\!=\!1) P(Y\!\mid\!\doo(A), L_r, R_K,R_{M_r}\!=\!1)$. Given assumption \textit{(ii)}, the first term on the right-hand side drops the intervention $\doo(A)$. By assumption \textit{(iv)} and Rule 1, the second term on the right-hand side becomes $P(Y\!\mid\!\doo(A), L_r, R_K,R_{M_r}\!=\!1,R_H\!=\!0)$. Finally, given condition on context $R_K\!=\!r,R_H\!=\!0$, analysis of such last term can be done in the graph that removes from $\mathcal{G}_r$ edges labeled with ``$R_H \!=\! 0$''. Assumptions \textit{(ii)} and \textit{(v)}, and Rule 2 of \textit{do}-calculus, applied to the last term lead directly to \cref{eq:nate1,eq:nate2}. \hfill$\square$

\Cref{eq:nate} expresses the NATE as a weighted average of context CATEs, denoted $\Delta_a\vartheta_r(a)$, where the weights correspond to the marginal probability of a missingness pattern on $R_K$ occurring. 




In the scenario depicted in \cref{fig1c}, one can set $K=Y_0$, $H=\emptyset$, $L_0=W$ and $L_1=\{W,Y_0\}$. The graph $\mathcal{G}_0=\mathcal{H}_0$ removes the arrows $ \{(Y_0, A) , (Y_0, Y_1) \}$. Therefore, in the context $R_{Y_0} = 0$, $P(Y \mid \doo(A), R_{Y_0} = 0)$ is recovered from $ \mathcal{G}_0$ as:
\begin{equation*}
    \int \dd P(W \mid R_{Y_0} = 0)\, P(Y \mid W, A, R_{Y_0} = 0).
\end{equation*}

In addition, the graph $\mathcal{G}_1=\mathcal{H}_1 $ retains all the arrows. Thus, in the context $R_{Y_0} = 1$, $P(Y \mid \doo(A), R_{Y_0} = 1) $ is recovered from $ \mathcal{G}_1$ as:  
\begin{equation*}
\int \dd P(W, Y_0 \mid R_{Y_0} = 1)\, P(Y \mid W, Y_0, A, R_{Y_0} = 1).
\end{equation*}

Consequently, the NATE is given by:
\begin{align}\label{eq:zate1}
& \theta = \mathbb{P}(R_{Y_0}=0)\, \E_{W\mid R_{Y_0}=0} \Delta_a Q_0(W,a)\\ \notag
&\quad + \mathbb{P}(R_{Y_0}=1)\, \E_{W,Y_0\mid R_{Y_0}=1}  \Delta_a Q_1(W,Y_0,a), \\ \label{eq:zate2}
& Q_0(W,A) = \E \left[Y\mid W,A=a,R_{Y_0}=0 \right],
\end{align}

with $Q_1(W,Y_0,A)$ given in \cref{eq:rateeq2}.

In the scenario in \cref{fig3c}, the NATE is recovered under \cref{prop:4} by setting $K=\emptyset$, which makes all $L$-sets empty by convention; and $H=W$, which highlights the relevant role of such set.

The recovered estimands for the FATE, in \cref{eq:rateeq1}, and the NATE, in \cref{eq:zate1}, highlight two key differences between these queries. The FATE relies on a single common response function, $ Q_1 $, taking $ Y_0 $ as input. This function can be learned from units with observed $ Y_0 $, and then imputed for the remaining units. In contrast, the NATE requires two separate response functions, adding $Q_0$, which does not use $Y_0$ as input. Each of these estimands can be learned from their respective subpopulation, weighted by their size, and averaged, without any imputation.

The NATE and FATE estimands may be motivated by distinct policy designs. For instance, if a proposed policy requires complete observability of certain variables for all units ---such as mandatory pre-testing--- then the FATE becomes a more relevant target for pre-evaluating its effectiveness, as it reflects potential outcomes under a scenario of full information. In contrast, the NATE is more appropriate when such observability is not enforced, but missingness remains prevalent. In these cases, the NATE captures the causal effect accounting for mechanism shifts due to missing data and better reflects the anticipated performance of the policy under conditions where not all inputs are observed.

\section{Estimation}\label{sec:estim}

Building upon the foundational work on doubly-robust estimation of causal effects in settings with missing data \parencite{Bang,AIPW2017}, a surge of recent research has lead to the development of misspecification-robust estimation methods, particularly under missing outcome data \parencite{Wei2022, multiplyDR, dw, Tang2024, Huber, ours}. Mechanism shifts can have significant implications for the efficient estimation of causal queries, especially for robust estimation in a semiparametric model. Doubly-robust estimators for causal parameters remain consistent provided that at least one of their components ---either the outcome regression or the treatment assignment model--- is correctly specified. However, under CSI either of these components may be affected by shifts. 

When missingness-specific independencies are relevant, consistent and robust estimation of the NATE requires tailored parametrization of the nuisance components. Although one could estimate each context CATE separately and compute a weighted average based on context frequencies, this approach may suffer for contexts with sparse observations. In such cases, pooling information from and across more populous contexts can enhance estimation quality. 

Building upon the example introduced in the motivation and illustrated in \cref{fig1c}, we now present a doubly-robust estimator for the NATE. We parameterize the outcome models $Q_1$, in \cref{eq:rateeq2}, and $Q_0$, in \cref{eq:zate2}, jointly as:
\begin{equation*}
    Q_r(W,Y_0,A) := Q_0(W,A) + r\cdot q(W,Y_0,A),
\end{equation*}

with $q$ being an auxiliary function. Since this $lm$-graph implies that the exposure mechanism is also shifted by the missingness of $Y_0$, this dependence should be reflected in the parametrization of the propensity score, specified as:
\begin{equation*}
    \log\frac{\pi_r(W,Y_0)}{1-\pi_r(W,Y_0)} = \log\frac{\pi_0(W)}{1-\pi_0(W)} + r\cdot \rho(W,Y_0),
\end{equation*}

where $\pi_1(W,Y_0):=\mathbb{P}(A=1\mid W,Y_0)$ and $\pi_0(W):=\mathbb{P}(A=1\mid W)$  under common positivity constraints, i.e., $0<\pi_0(W),\pi_1(W,Y_0)<1$ (almost surely), and employing an auxiliary function $\rho$.

Then, a 
\textit{one-step corrected} estimator of the NATE, using data $\{(W_i,Y_{0,i}^\dagger, R_{Y_{0},i}=r_i, A_i, Y_{1,i})\}_i^n$, is given by
\begin{align}\notag
    & \widehat{\theta} = \frac{1}{n}\sum_{i=1}^n \Delta_a \widehat{Q}_{r_i}(W_i,Y_{0,i},a)\ +  \\ \label{eq:drn} 
    &\!\! \frac{A_i-\widehat{\pi}_{r_i}(W_i,Y_{0,i})}{\widehat{\pi}_{r_i}(W_i,Y_{0,i})[1-\widehat{\pi}_{r_i}(W_i,Y_{0,i})]}\left[Y_{1,i}- \widehat{Q}_{r_i}(W_i,Y_{0,i},A_i)\right]
\end{align}

Intuitively, this corresponds to the conventional AIPW estimator \parencite{robins1995analysis, robins1997comment}, where the outcome and propensity score models, $Q_r$ and $\pi_r$, are parameterized to account for different adjustment sets based on the context $R_{Y_0}=r$. Unfortunately, in more general settings where $\abs{\mathcal{R}_{\rm{sh}}}$ is large, a combinatorial explosion of nuisance parameters arises, making it challenging to determine optimal pooling parametrizations or constraints. This, in turn, complicates the computation by hand of the required influence functions and efficient estimators.

\section{Simulations and Application}\label{sec:simu}

\begin{figure}[t]
\centering
\begin{subfigure}{.2275\textwidth}
  \centering
\begin{tikzpicture}[scale=0.83, transform shape]
  \begin{axis}
    [   width=5.5cm, 
        height=4.5cm,  
        ymin=-6, ymax=4.5,  
    boxplot/draw direction=y,
    xtick={1,2,3,4}, ytick={-6,-4,-2,0,2,4},
    xticklabels={DR.F, DR.N, Imp., MIM},
    ylabel = {Estimate causal effect},
    x tick label style={font=\footnotesize, text width=1.0cm, align=center}
    ]
    \addplot+[mark = *, mark options = {red!50},
    boxplot prepared={
        lower whisker = -5.178008,
        lower quartile = -4.382609,
        median = -4.018687,
        upper quartile = -3.690454,
        upper whisker = -2.694362,
        average = -4.017672,
    }, color = red!100, thick
    ] coordinates{};
    \addplot+[mark = *,mark options = {blue!50},
    boxplot prepared={
        lower whisker = 0.4144086,
        lower quartile = 1.0272653,
        median = 1.2753270,
        upper quartile = 1.5011448,
        upper whisker = 2.0350559,
        average = 1.262105,
    }, color = blue!100, thick
    ] coordinates{};
    \addplot+[mark = *,mark options = {black!50},
    boxplot prepared={
        lower whisker = 2.274969,
        lower quartile = 2.867985,
        median = 3.083811,
        upper quartile = 3.292128,
        upper whisker = 3.923265,
        average = 3.079754,
    }, color = black!100, thick
    ] coordinates{};
    \addplot+[mark = *,mark options = {black!50},
    boxplot prepared={
        lower whisker = 1.170960,
        lower quartile = 1.567742,
        median = 1.743504,
        upper quartile = 1.924765,
        upper whisker = 2.447416,
        average = 1.741383,
    }, color = black!100, thick
    ] coordinates{};
    \draw[thick, dashed, red] (0,-4.010287) -- (5,-4.010287);
    \draw[thick, dashed, blue] (0,1.261823) -- (5,1.261823);
g    \end{axis}
\end{tikzpicture}
    \caption{}
    \label{fig4a}
\end{subfigure}%
\begin{subfigure}{.2275\textwidth}
  \centering
\begin{tikzpicture}[scale=0.83, transform shape]
  \begin{axis}
    [   width=5.5cm, 
        height=4.5cm,  
        ymin=-6, ymax=4.5,  
    boxplot/draw direction=y,
    xtick={1,2,3,4}, ytick={-6,-4,-2,0,2,4},
    xticklabels={DR.F, DR.N, Imp., MIM},
    x tick label style={font=\footnotesize, text width=1.0cm, align=center},
    yticklabels={},
    y tick label style={draw=none},
    ]
    \addplot+[mark = *, mark options = {red!50},
    boxplot prepared={
        lower whisker = -5.307663,
        lower quartile = -4.378740,
        median = -4.023757,
        upper quartile = -3.723088,
        upper whisker = -2.803664,
        average = -4.024713,
    }, color = red!100, thick
    ] coordinates{};
    \addplot+[mark = *,mark options = {blue!50},
    boxplot prepared={
        lower whisker = -1.69007134,
        lower quartile = -1.11406314,
        median = -0.88523144,
        upper quartile = -0.64939930,
        upper whisker = -0.04761483,
        average = -0.8773132,
    }, color = blue!100, thick
    ] coordinates{};
    \addplot+[mark = *,mark options = {black!50},
    boxplot prepared={
        lower whisker = -0.1964109,
        lower quartile = 0.2131180,
        median = 0.3598920,
        upper quartile = 0.5939617,
        upper whisker = 1.1088804,
        average = 0.3968317,
    }, color = black!100, thick
    ] coordinates{};
    \addplot+[mark = *,mark options = {black!50},
    boxplot prepared={
       lower whisker = -1.2114910,
        lower quartile = -0.7458321,
        median = -0.5647480,
        upper quartile = -0.3808295,
        upper whisker = 0.1384041,
        average = -0.5541139,
    }, color = black!100, thick
    ] coordinates{};
    \draw[thick, dashed, red] (0,-4.007316) -- (5,-4.007316);
    \draw[thick, dashed, blue] (0,-0.8844339) -- (5,-0.8844339);
    \end{axis}
\end{tikzpicture}
    \caption{}
    \label{fig4c}
\end{subfigure}%
\caption{Boxplots of sampling distribution of estimators in simulation exercise with sample size $ n = 5\,000 $ and $ m = 200 $ repetitions, under missingness rates of (a) 50\% and (b) 30\%. Broken lines indicate the oracle FATE (in \textcolor{red}{red}) and NATE (in \textcolor{blue}{blue}). Abbr.: DR.F = doubly-robust FATE estimator (\cref{appendixA}), DR.N = doubly-robust NATE estimator (\cref{sec:estim}), Imp = Multiple Imputation Method, MIM = Missing Indicator Method.}
\label{fig4}
\end{figure}
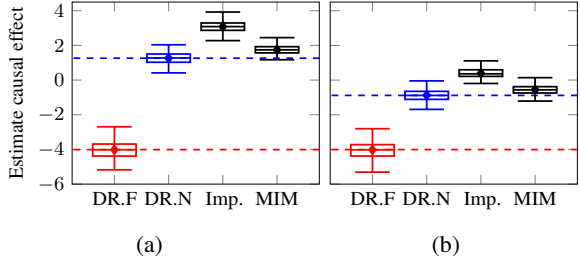

We conduct a simulation study to compare estimators for the FATE and NATE for the case of $lm$-graph in \cref{fig1c}. Data are generated from an $lm$-SCM with varying parameterizations for two scenarios on the missingness rate of $Y_0$: (a) 50\% and (b) 30\%. The FATE is estimated using a doubly-robust method (DR.F) that imputes pseudo-outcomes via one-step corrections from \cref{eq:rateeq1}. The NATE is estimated using the doubly-robust estimator (DR.N) defined in \cref{eq:drn}. We also include a Multiple Imputation estimator (Imp.) and a Missing Indicator Method (MIM) estimator, which uses $(W, A, R_{Y_0} \cdot Y_0)$ as covariates and includes interactions of $W,A$ with $R_{Y_0}$ \parencite{greenland1995critical, jones1996indicator}. Implementation details are in \cref{appendixA}.

Results in \cref{fig4} display the sampling distribution of the estimators with sample size $n = 5\,000$ and $m = 200$ repetitions. They highlight key differences between the FATE and NATE, which may occasionally have opposite signs, though the difference depends on the missingness rate. The DR.N estimator performs well across the two setups. Under data generated from an $lm$-SCM, both Imp. and MIM estimators target the NATE. The Imp. estimator is biased, as its imputation model fails to fully capture mechanism shifts, while MIM shows less bias, supporting its empirical value \parencite{song2021, missingCov}, when justified graphically and implemented with flexible models. 



As a motivating application, we study the impact of pharmacological treatment for \textit{attention-deficit/hyperactivity disorder} (ADHD) upon national numeracy test scores among Norwegian 8th-grade students diagnosed with ADHD, using observational data. A key confounder, the grade 5 test score, is missing in 10\% of the 8\,450 cases due to both exogenous and endogenous factors. Further details are given in \cref{appendixB}. The results in \cref{fig5} reveal a small effect of medication on scores. One possible explanation for the small positive difference between the NATE and FATE estimates is a compensatory or diminishing returns effect: prior testing-experience also improves performance, and medication may provide greater benefits to children without such experience. Thus, when test-taking in grade 5 is made mandatory, the added impact of medication is slightly reduced. Latent factors could also explain such difference.

\begin{figure}[t]
  \centering
\begin{tikzpicture}
    \begin{axis}[
    scale=0.85, transform shape,
        width=5.5cm, 
        height=4.0cm,  
      xmax=1.9,
      xmin=0.3,
      ymin=0,
      ymax=6,
      ytick={1,...,5},
      axis y line*=right,
      axis x line*=bottom,
      yticklabels={DR.F, DR.N, MIM, Imp., CC.},
      xlabel={raw score points},
      y tick label style={font=\scriptsize,text width=1.0cm},
      x tick label style={font=\scriptsize,align=center}]
    \addplot+ [red, solid, style=thick, boxplot prepared={
        lower whisker=0.4734042, median=0.8782181, upper whisker=1.283032
        }] coordinates {}; 
    
    \addplot+ [blue, solid, style=thick, boxplot prepared={
        lower whisker=0.6259206, median=1.01899, upper whisker=1.412059
        }] coordinates {}; 
    
    \addplot+ [black, solid, style=thick, boxplot prepared={
        lower whisker=0.7525132, median=1.150372, upper whisker=1.548231
        }] coordinates {};

    \addplot+ [black, solid, style=thick, boxplot prepared={
        lower whisker=0.8609987432665009, median=1.1663839728113374, upper whisker=1.471769202356174
        }] coordinates {}; 

     \addplot+ [gray, solid, style=thick, boxplot prepared={
        lower whisker=1.06959477046, median=1.37498, upper whisker=1.68036522954
        }] coordinates {}; 
    
    \addplot[color=black,dashed] coordinates {
        		(0,6)
        		(0,0)
        	};
    \end{axis}
    \begin{axis}[
    scale=0.85, transform shape,
        width=5.5cm, 
        height=4.0cm,  
      xmax=1.9/11,
      xmin=0.3/11,
      ymin=0,
      ymax=6,
      ytick={1,...,5},
      axis y line*=left,
      axis x line*=top,
      yticklabels={,,},
      xlabel={Cohen's $d$}, 
      y tick label style={font=\scriptsize,align=center},
      x tick label style={font=\scriptsize,align=center}
      ]
      \end{axis}
\end{tikzpicture}%
\caption{Estimated effects and confidence intervals for different estimators in the application case. Abbr.: CC = Complete cases.}
\label{fig5}
\end{figure}
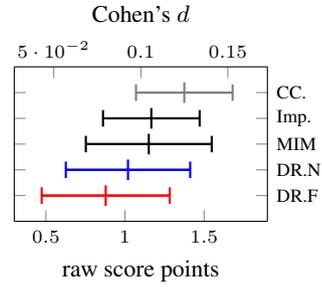

\section{Conclusion and Discussion}\label{sec:conclusion}

We examined the problem of mechanism shifts that cannot be captured by existing $m$-graph-based frameworks for causal inference with missing data. To address these limitations, we introduced $lm$-SCMs and $lm$-graphs, which retain the expressiveness of $m$-graphs while incorporating a set of CSIs. We defined relevant causal effects and established recovery criteria for them. Given that causal effect identification in systems with CSI is generally NP-hard, our recovery criteria for the \textit{natural} effects are only declarative and sufficient. Future research will further investigate other causal queries within $lm$-SCMs with categorial or continuous exposures, broad recovery criteria, and the issue of testability implications and impossibilities under $lm$-graphs.

\clearpage




\onecolumn
\begin{multicols}{2}

\begin{acknowledgements} 

We thank the organizing committee of the European Causal Inference Meeting (2025) in Ghent for the opportunity to present an early version of this work and for the valuable feedback received. This research was supported by the Research Council of Norway (project number 301081, PI: Guido Biele) and approved by the Regional Committees for Medical and Health Research Ethics (REK) under approval number 96604. Johan Pensar received additional support from the Research Council of Norway through INTEGREAT -- The Norwegian Centre for Knowledge-Driven Machine Learning (project number 332645).

\end{acknowledgements}

\sloppy
\printbibliography{}
\end{multicols}


\clearpage
\appendix


\title{Causal Inference amid Missingness-Specific independencies\\ and Mechanism Shifts (Supplementary Material)}
\maketitle

\section{Simulation task details}\label{appendixA}

For each sampling batch, we generate 5\,000 i.i.d. samples from the following SCM, with associated $lm$-graph in \cref{fig1c}:
\begin{align}
W &\sim N(0,1), &\\
Y_0 &= -3-2\,W+W^2 + U_0, \quad & U_0\sim N(0,7),&\\
R_{Y_0} &\sim \operatorname{Ber}_{\operatorname{logit}}(\beta_1 + \beta_2\,W), &\\
A &\sim \operatorname{Ber}_{\operatorname{logit}}\left(R_{Y_0}(W+0.3\,Y_{0})+(1-R_{Y_0})(-0.5 + 1.5\,W) \right), &\\
Y &=  R_{Y_0}(3 + 1.8\,W - 2\,A -1.5\,Y_{0} - 0.8\,A\,W + 4\,A\,Y_{0})   & \\
 &\qquad + (1-R_{Y_0})(4+ 6\,W + 8\,A -8 \,W\,A ) + U_1,\quad & U_1\sim N(0,7).
\end{align}

Here, $N(0,\sigma) $ denotes a Gaussian distribution with mean zero and standard deviation $\sigma $, while $\operatorname{Ber}_{\operatorname{logit}}(L) $ represents a Bernoulli distribution parameterized via the logit function, meaning the success probability is given by $p = (1+\exp(-L))^{-1} $. Varying the parameters $\beta_1 $ and $\beta_2 $ leads to different missingness rates. Specifically, when $\beta_1 = -0.2 $ and $\beta_2 = -1.2 $, it is around 50\%; and for $\beta_1 = 1.1 $ and $\beta_2 = -1.0 $, it is approximately 30\%.

The doubly-robust estimator for the FATE (DR.F) was computed using  \textit{one-step corrected} pseudo-outcomes, akin to the approach propsed by \textcite{DRlearner} to estimate the CATE. Let the regression surface $\widehat{Q}_1(W,Y_0,A)= \widehat{\E}\left[Y\mid W,Y_0,A,R_{Y_0}=1\right]$ be learned from units with $R_{Y_0}=1$. The pseudo-outcomes $\widetilde{\delta}(W,Y_0,A)$ are given by:
\begin{equation*}
    \widetilde{\delta}(W,Y_0,A) := \Delta_a \widehat{Q}_1(W,Y_0,a) + \frac{A-\widehat{\pi}_{1}(W,Y_{0})}{\widehat{\pi}_{1}(W,Y_{0})[1-\widehat{\pi}_{r_i}(W,Y_{0})]}\left[Y_{1}- \widehat{Q}_{1}(W,Y_{0},A)\right].
\end{equation*}

And $\widehat{\tau}(W)$ are the predictions from the meta-regression model $\widehat{\tau}(W)= \widehat{\E}\left[\widetilde{\delta}(W,Y_0,A)\mid W,R_{Y_0}=1\right]$ across all units. The DR.F estimator is then given by:
\begin{equation*}
    \hat{\phi}=\frac{1}{n}\sum_{i=1}^n\left[ \widehat{\tau}(W_i)+\frac{R_{Y_0,i}}{\widehat{\mathbb{P}}(R_{Y_0}=1\mid W_i)}(\widehat{\tau}(W_i)- \Delta_a \widehat{Q}_1(W_i,Y_{0,i},a))\right].
\end{equation*}

An estimate of its variance, to construct confidence intervals, can be obtained either through bootstrap methods or via an estimate of the asymptotic variance given by: 
\begin{equation*}
    \frac{1}{n^2}\sum_{i=1}^n\left[ \widehat{\tau}(W_i)+\frac{R_{Y_0,i}}{\widehat{\mathbb{P}}(R_{Y_0}=1\mid W_i)}(\widehat{\tau}(W_i)- \Delta_a \widehat{Q}_1(W_i,Y_{0,i},a))- \hat{\phi}\right]^2.
\end{equation*}

Both the DR.N and DR.F estimators exhibit desirable asymptotic properties under several technical conditions. These include: \textit{(i)} a fully nonparametric or saturated model for the data-generating process, \textit{(ii)} smoothness of model paths, \textit{(iii)} positivity of the relevant propensity scores, \textit{(iv)} boundedness of the outcome mean, Gâteaux derivative, and their variances, \textit{(v)} the Donsker class condition (i.e., bounded estimator complexity), and \textit{(vi)} sufficiently fast convergence of nuisance estimates and their interaction terms. For further details on one-step corrections, we refer the reader to \textcite{hines2022demystifying}.

The data and code utilized for the simulation study are accessible in a personal GitHub repository at \url{https://github.com/johandh2o}. 

\section{Application case details}\label{appendixB}

We assess the impact of pharmacological treatment with stimulant medication upon the numeracy test scores at grade 8 obtained by Norwegian children diagnosed with ADHD. By integrating information from national registries, we compile data on the medication history and national test scores of all children diagnosed with ADHD born between 2000 and 2007 in Norway, who would go to take the national test up to 2021. We exclude those with severe comorbid disorders (totaling $8\,450$ individuals). Variables at the student, family, and school levels are linked from the Norwegian Prescription Database (NorPD), the Norwegian Patient Registry (NPR), the Database for Control and Payment of Health Reimbursement (KUHR), Statistics Norway (SSB), and the Medical Birth Registry of Norway (MBRN). We leverage data on students' and parents' diagnoses and their consultations with medical services during the pre-exposure period. To operationalize relevant variables, we employ the following grouping:
\begin{itemize}
    \item \textbf{Pre-exposure covariates} $W$: sex at birth, birth year/month cohorts, birth parity number, raw score at grade 5 national test for numeracy, mother's education level, mother's age at birth, student's and parents' diagnoses and medical consultations for related comorbid disorders, school identification (fixed effect), prior dispensations of ADHD stimulant medication for at least 90 days, and duration of prior treatment.
    \item \textbf{Exposure} $A$: 
    having received dispensations of ADHD stimulant medication for at least 75\% of the prescribed treatment period between the start of grade 6 and the national test in grade 8.
    \item \textbf{Outcomes} $Y$: raw scores at grade 8 national test for numeracy.
\end{itemize}

The missingness rate on test scores at grade 5 is about 10\%. This rate is not expected to induce a disproportionate amount of selection bias but, from a public health standpoint, it could still lead to shifts in conclusions and policy design that may not be fully aligned with the target population. All estimators were fitted with sample-splitting and super-learning schemes, with a battery of four base algorithms: generalized linear models (GLM), penalized GLMs, random forest (RF), and boosted decision trees (BDT).

Due to the sensitive nature of the research topic, data from the application case cannot be obtained from the authors, but have to be requested from the relevant Norwegian authorities.

\end{document}